\documentclass[letterpaper, superscriptaddress, twocolumn, showpacs,
nofootinbib]{revtex4}

\usepackage[]{graphicx}
\usepackage{amsmath}
\usepackage{amsfonts}
\usepackage{color}
\usepackage{hyperref}
\usepackage{bm}
\usepackage{graphicx}
\usepackage{amsbsy}

\def\ket#1{| #1 \rangle}
\def\bra#1{\langle #1 |}

\def\bk#1#2{\langle #1 | #2 \rangle}

\newcommand{\aZ}{{\widetilde{Z}}}

\newcommand{\E}{{\mathcal{E}}}
\newcommand{\F}{{\mathcal{F}}}
\newcommand{\1}{{\openone}}

\newcommand{\erf}{{\text{erf}}}
\newcommand{\erfc}{{\text{erfc}}}

\pacs{03.67.Lx, 03.67.Mn, 42.65.-k}
 
\date{\today}

\begin{document}

\title{Computation with Coherent States via Teleportations to and from a Quantum Bus}

\author{Marcus Silva}
\affiliation{Department of Physics and Astronomy, and Institute for Quantum Computing, University of
Waterloo, ON, N2L 3G1, Canada}
\author{Casey R.~Myers}
\affiliation{ National Institute of Informatics, 2-1-2 Hitotsubashi,
Chiyoda-ku, Tokyo 101-8430, Japan}
\affiliation{Department of Physics and Astronomy, and Institute for Quantum Computing, University of
Waterloo, ON, N2L 3G1, Canada}

\begin{abstract}
In this paper we present results illustrating the power and flexibility of one-bit teleportations in quantum bus computation. We first show a scheme to perform a universal set of gates on continuous variable modes, which we call a quantum bus or {\em qubus}, using controlled phase-space rotations, homodyne detection, ancilla qubits and single qubit measurement. The resource usage for this scheme is lower than any previous scheme to date. We then illustrate how one-bit teleportations into a qubus can be used to encode qubit states into a quantum repetition code, which in turn can be used as an efficient method for producing GHZ states that can be used to create large cluster states. Each of these schemes can be modified so that teleportation measurements are post-selected to yield outputs with higher fidelity, without changing the physical parameters of the system.
\end{abstract}

\maketitle

\section{\label{sec:intro} Introduction}

Bennett~{\em et~al.}~\cite{Bennett93} showed that an unknown quantum state (a qubit) could be {\em teleported} via two classical bits with the use of a maximally entangled Bell state shared between the sender and receiver.  The significance of teleportation as a tool for quantum information was extended when Gottesman and Chuang~\cite{Gottesman99} showed that unitary gates could be performed using modified teleportation protocols, known as {\em gate teleportation}, where the task of applying a certain gate was effectively translated to the task of preparing a certain state. Since then teleportation has been an invaluable tool for the quantum information community, as gate teleportation was the basis for showing that linear optics with single photons and photo-detectors was sufficient for a scalable quantum computer~\cite{KLM}. Moreover, Zhou {\em et~al.}~\cite{Zhou00} demonstrated that all previously known fault-tolerant gate constructions were equivalent to {\em one-bit teleportations} of gates.

Recently, the use of one-bit teleportations between a qubit and a continuous variable {\em quantum bus} (or {\em qubus}) has been shown to be important for fault-tolerance~\cite{StabPap}. Using one-bit teleportations to transfer between two different forms of quantum logic, a fault tolerant method to measure the syndromes for any stabiliser code with the qubus architecture was shown, allowing for a linear saving in resources compared to a general CNOT construction. In terms of optics, the two different types of quantum logic used were polarisation $\{\ket{0}=\ket{H},\ket{1}=\ket{V}\}$ and logical states corresponding to rotated coherent states $\{\ket{\alpha},\ket{e^{\pm i\theta}\alpha}\}$, although in general any two-level system (qubit) which can interact with a continuous variable mode (qubus) would suffice. The relative ease with which single qubit operations can be generally performed prompted the question of whether a universal set of gates can be constructed with this rotated coherent state logic. In this paper we describe one such construction, which we call {\em qubus logic}.

The fault-tolerant error-correction scheme using a qubus~\cite{StabPap} exploits the fact that entanglement is easy to create with coherent cat states of the qubus, such as $\ket{\alpha}+\ket{\alpha e^{i\theta}}$, and single qubit operations are easily performed on a two-level system. In this paper we describe how these cat sates can be used as a resource to construct other large entangled states, such as cluster states~\cite{Raussendorf01,Nielsen04,Browne05}, using one-bit teleportations between a qubit and a qubus.

Although the average fidelities of qubus logic and cluster state preparation are dependent on how strong the interaction between the qubit and the qubus can be made, and how large the amplitude $\alpha$ is, these fidelities can be increased arbitrarily close to $1$ through the use of post-selection during the one-bit teleportations, demonstrating the power and flexibility of teleportation in qubus computation for state preparation. 

The paper is organised as follows. First, in Section ~\ref{sec:onebitteleportations} we revisit one-bit teleportations for the qubus scheme. Next, in Section ~\ref{sec:coherentstatelogic} we present a technique to perform quantum computation using coherent states of the qubus as basis states.  To do this we make use of controlled (phase-space) rotations and ancilla qubits. This coherent state computation scheme is the most efficient to date. In Section ~\ref{sec:clusterstate} we show how we can efficiently prepare repetition encoded states using one-bit teleportations, and how such encoders can be used to prepare large cluster states.

\section{\label{sec:onebitteleportations}One-Bit Teleportations}

In the original quantum teleportation protocol an arbitrary quantum state can be transferred between two parties that share a maximally entangled state by using only measurements and communication of measurement outcomes~\cite{Bennett93}. Modifications of the resource state allow for the applications of unitaries to an arbitrary state in a similar manner, in what is known as {\em gate teleportation}~\cite{Gottesman99}. The main advantage of gate teleportation is the fact that it allows for the application of the unitary to be delegated to the state preparation stage. In some physical realisations of quantum devices, it may only be possible to prepare these states with some probability of success. In that case, the successful preparations can still be used for scalable quantum computation~\cite{Gottesman99}. When dealing with noisy quantum devices, it is important to encode the quantum state across multiple subsystems, at the cost of requiring more complex operations to implement encoded unitaries. In order to avoid the the uncontrolled propagation of errors during these operations, one can also employ gate teleportation with the extra step of verifying the integrity of the resource state before use~\cite{Gottesman99,Zhou00,KLM,Knill,SDKO}. In the cases where the teleportation protocol is used only to separate the preparation of complex resource states from the rest of the computation, simpler protocols can be devised. These protocols are known as {\em one-bit teleportations}~\cite{Zhou00}. Unitaries implemented through one-bit gate teleportation can also be used for fault-tolerant quantum computation~\cite{Zhou00} as well as measurement based quantum computation~\cite{Raussendorf01}. The main difference between one-bit teleportation and the standard teleportation protocol is the lack of a maximally entangled state. Instead, in order to perform a one-bit teleportation it is necessary that the two parties interact directly in a specified manner, and that the qubit which will receive the teleported state be prepared in a special state initially.

\begin{figure}[ht]
\includegraphics[width=8cm]{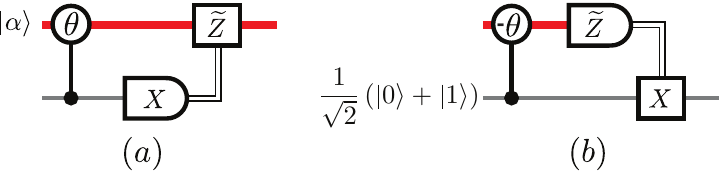}
\caption{\footnotesize Approximate one-bit teleportation 
protocols~\cite{Zhou00} using controlled rotations. Here, the light grey
lines correspond to qubits, and the thick red lines correspond to quantum
bus modes.} 
\label{one-teleWNL}
\end{figure}

Some unitary operations on coherent states can be difficult to implement deterministically, while the creation of entangled multimode coherent states is relatively easy. Single qubits, on the other hand, are usually relatively easy to manipulate, while interactions between them can be challenging. For this reason, we consider one-bit teleportation between states of a qubit and states of a field in a quantum bus, or {\em qubus}. The two types of one-bit teleportations for qubus computation are shown in Fig.~(\ref{one-teleWNL}), based on similar constructions proposed for qubits by Zhou {\em et al.}\cite{Zhou00}. 

The one-bit teleportation of the qubit state $a\ket{0}+b\ket{1}$ into the state of the qubus, in the coherent state basis $\{\ket{\alpha},\ket{\alpha e^{i\theta}}\}$, is depicted in Fig.~(\ref{one-teleWNL}a). The qubit itself can be encoded, for example, in the polarisation of a photon, i.e. $\ket{0}=\ket{H}$ and $\ket{1}=\ket{V}$. The initial state, before any operation, is $\bigl(a\ket{0}+b\ket{1}\bigr)\ket{\alpha}$. The controlled phase-space rotation corresponds to the unitary which applies a phase shift of $\theta$ to the bus if the qubit state is $\ket{1}$, and does nothing otherwise~\footnote{This can be implemented by an interaction of the Jaynes-Cummings type between the qubit and the qubus, in the dispersive limit.}. After the controlled rotation by $\theta$ the state becomes $a\ket{0}\ket{\alpha}+b\ket{1}\ket{e^{i\theta}\alpha}$. Representing the qubit state in the Pauli $X$ eigenbasis, this is $\ket{+}\bigl(a\ket{\alpha}+b\ket{e^{i\theta}\alpha}\bigr)/\sqrt{2}+\ket{-}\bigl(a\ket{\alpha}-b\ket{e^{i\theta}\alpha}\bigr)/\sqrt{2}$. When we detect $\ket{+}$ we have successfully teleported our qubit into $\ket{\alpha}$, $\ket{e^{i\theta}\alpha}$ logic.  When we detect $\ket{-}$ we have the state $a\ket{\alpha}-b\ket{e^{i\theta}\alpha}$. The relative phase discrepancy can be corrected by the operation $\tilde{Z}$, which approximates the Pauli $Z$ operation in the $\{\ket{\alpha},\ket{\alpha e^{i\theta}}\}$ basis. This correction can be delayed until the state is teleported back to a qubit, where it is more easily implemented.

The one-bit teleportation of the state $a\ket{\alpha}+b\ket{\alpha e^{i\theta}}$ of the qubus to the state of the qubit can be performed by the circuit depicted in Fig.~(\ref{one-teleWNL}b). That is, we start with the state $\bigl(a\ket{\alpha}+b\ket{\alpha e^{i\theta}}\bigr)(\ket{0}+\ket{1})/\sqrt{2}$. After the controlled rotation by $-\theta$, the state becomes $\ket{\alpha}\bigl( a\ket{0}+b\ket{1}\bigr)/\sqrt{2}+\bigl(b\ket{e^{i\theta}\alpha}\ket{0}+a\ket{e^{-i\theta}\alpha}\ket{1}\bigr)/\sqrt{2}$. Projecting the qubus state into the $x$-quadrature eigenstate $\ket{x}$ via homodyne detection, which is the measurement we depict as $\aZ$, we obtain the conditional unnormalised state $\ket{\psi(x)}$ 
\begin{multline}\label{busbitteleport}
\ket{\psi(x)} = \frac{f(x,\alpha)}{\sqrt{2}}(a\ket{0}+b\ket{1}) \\ 
 + \frac{f(x,\alpha\cos(\theta))}{\sqrt{2}}(e^{i\phi(x)}b\ket{0}+e^{-i\phi(x)}a\ket{1})
\end{multline}
where
\begin{gather}
f(x,\beta) = \frac{1}{(2\pi)^4}\exp\left(\frac{-(x-2\beta)^2}{4}\right)\\
\phi(x) = \alpha x \sin(\theta)-\alpha^2\sin(2\theta),
\end{gather}
since $\langle x | \alpha e^{\pm i \theta}\rangle = e^{\pm i \phi(x)}f(x,\alpha\cos(\theta))$
and $\langle x | \alpha \rangle = f(x,\alpha)$ for real $\alpha$~\cite{Gardiner, Barret05}. 

The weights $f(x,\alpha)$ and $f(x,\alpha\cos(\theta))$ are Gaussian functions  with the same variance but different means, given by $2\alpha$ and $2\alpha\cos(\theta)$, respectively. Given $x_0=\alpha(1+\cos(\theta))$, the midpoint between $f(x,\alpha)$ and $f(x,\alpha\cos(\theta))$, one can maximise the fidelity of obtaining the desired state $a\ket{0}+b\ket{1}$ (averaged over all possible values of $x$) by simply doing nothing when $x>x_0$ (where $f(x,\alpha)>f(x,\alpha\cos(\theta))$), or applying $Z_{\phi(x)}=\exp(-i\phi(x)Z)$, a Pauli $Z$ rotation by $\phi(x)$, followed by a Pauli $X$, when $x\le x_0$. For simplicity, the teleportation corrections are not explicitly depicted in the circuit diagrams.

\subsection{Average fidelities}

In order to quantify the performance of the protocols just described, consider the {\em process fidelity}~\cite{Jamiolkowski72,Horodecki99,Gilchrist05}. The process fidelity between two quantum operations is obtained by computing the fidelity between states isomorphic to the processes under the Choi-Jamio{\l}kowski isomorphism. For example, in order to compare a quantum process $\E$ acting on a $D$ dimensional system to another quantum process $\F$ acting on the same system, we compute the fidelity between the states 
\begin{gather}
\ket{\E}=\1_{1}\otimes\E_{2}\left(\frac{1}{\sqrt{d}}\sum_{i=1}^D\ket{ii}_{12}\right)\\
\ket{\F}=\1_{1}\otimes\F_{2}\left(\frac{1}{\sqrt{d}}\sum_{i=1}^D\ket{ii}_{12}\right).
\end{gather}
In the case of single qubit processes, we just need to consider the action of the process on one of the qubits of the state $\frac{1}{\sqrt{2}}(\ket{00}\pm\ket{11})$. The operational meaning of the process fidelity is given by considering the projection of the first qubit into a particular state $a\ket{0}+b\ket{1}$. In this case the second qubit collapses into the state corresponding to the output of the process acting on the state $a\ket{0}+b\ket{1}$. Thus a high fidelity between $\ket{\E}$ and $\ket{\F}$ implies a high fidelity between the outputs of the $\E$ and $\F$.

Consider the state produced by the circuit in Fig.~(\ref{one-teleWNL}a)
\begin{equation}\label{bitbusentangled}
\ket{\psi_\pm}=\frac{1}{\sqrt{2}}(\ket{0,\alpha}\pm\ket{1,\alpha e^{i\theta}}),
\end{equation}
which depends on the qubit measurement outcome. As the relative phase is known, and the correction can be performed after the state is teleported back to a qubit, for each of the outcomes we can compare this state with the ideal state expected from the definition of the basis states for the qubus. This results in the process fidelity of $1$ for one-bit teleportation into the qubus.

\begin{figure}[htb]
\centering
\includegraphics[width=8cm]{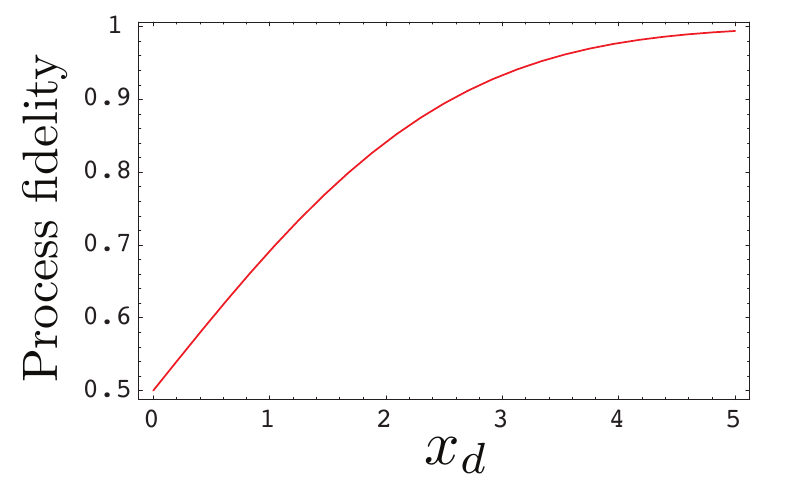}
\caption{\footnotesize Fidelity $F_p$ of one-bit teleportation from the qubus to a qubit, as a function of $x_d$.}
\label{onebitfidelity}
\end{figure}

For the case where we teleport the state from the qubus back into the qubit, using the circuit in Fig.~(\ref{one-teleWNL}b), we consider the action of the process on the second mode of the state $\ket{\psi_+}$ from Eq.~\eqref{bitbusentangled}. This is not, strictly speaking, the Choi-Jamio{\l}kowski isomorphism, but it gives the same operational meaning for the process fidelity as a precursor to the fidelity between the outputs of the different processes being compared, as any superposition of $\{\ket{\alpha},\ket{\alpha e^{i\theta}}\}$ can be prepared from $\ket{\psi_+}$ by projecting the qubit into some desired state. We expect the output state to be $\frac{1}{\sqrt{2}}\left(\ket{00}+\ket{11}\right)$ from the definition of the basis states, but we instead obtain the unnormalised states
\begin{multline}
\ket{\psi_E(x>x_0)} = \frac{f(x,\alpha)}{\sqrt{2}} \left(\frac{\ket{00}+\ket{11}}{\sqrt{2}}\right) + \\
\frac{f(x,\alpha\cos(\theta))}{\sqrt{2}} \left(\frac{e^{-i\phi(x)}\ket{01}+e^{i\phi(x)}\ket{10}}{\sqrt{2}}\right),
\end{multline}
\begin{multline}
\ket{\psi_E(x<x_0)} = \frac{f(x,\alpha)}{\sqrt{2}} \left(\frac{e^{-i\phi(x)}\ket{01}+e^{i\phi(x)}\ket{10}}{\sqrt{2}}\right) + \\\frac{f(x,\alpha\cos(\theta))}{\sqrt{2}} \left(\frac{\ket{00}+\ket{11}}{\sqrt{2}}\right).
\end{multline}
The normalised output state, averaged over all $x$ outcomes, is
\begin{multline}
\rho = \int_{x_0}^\infty \ket{\psi_E(x>x_0)} \bra{\psi_E(x>x_0)} dx + \\
\int_{-\infty}^{x_0} \ket{\psi_E(x<x_0)} \bra{\psi_E(x<x_0)} dx,
\end{multline}
so that the average process fidelity for one-bit teleportation into a qubit is 
\begin{equation}
F_p = \frac{1}{2} + \frac{1}{2}\erf\left(\frac{x_d}{2\sqrt{2}}\right)\label{eqn:SingeTeleF},
\end{equation}
where $x_d=2\alpha(1-\cos(\theta))\approx \alpha \theta^{2}$ for small $\theta$. 
Teleportation from the qubus into the qubit is not perfect, even in the ideal setting we consider, because the states $\ket{\alpha}$ and $\ket{e^{i \theta}\alpha}$ cannot be distinguished perfectly. However,  $F_p$ can be made arbitrarily close to one by letting $x_d\to\infty$, or $\alpha\theta^2\to\infty$ if $\theta\ll1$, as seen in Fig.~(\ref{onebitfidelity}). This corresponds to increasing the distinguishability of the coherent states $\ket{\alpha}$ and $\ket{e^{i \theta}\alpha}$.

\subsection{\label{sec:postselectedteleport}Post-selected teleportation}

In order to improve the average fidelity of the teleportations without changing the physical parameters $\alpha$ and $\theta$ of the basis states, one can post-select the outcomes of the $x$-quadrature measurements when teleporting states from the qubus mode to a qubit, as these outcomes essentially herald the fidelity of the output state with the desired state. Discarding the states with fidelity below a certain threshold allows for the average fidelity to be boosted, even in the case where $\alpha\theta^2\not\gg1$, at the cost of a certain probability of failure. This is particularly useful for the preparation of quantum states which are used as resources for some quantum information processing tasks.

\begin{figure}[htb]
\centering
\includegraphics{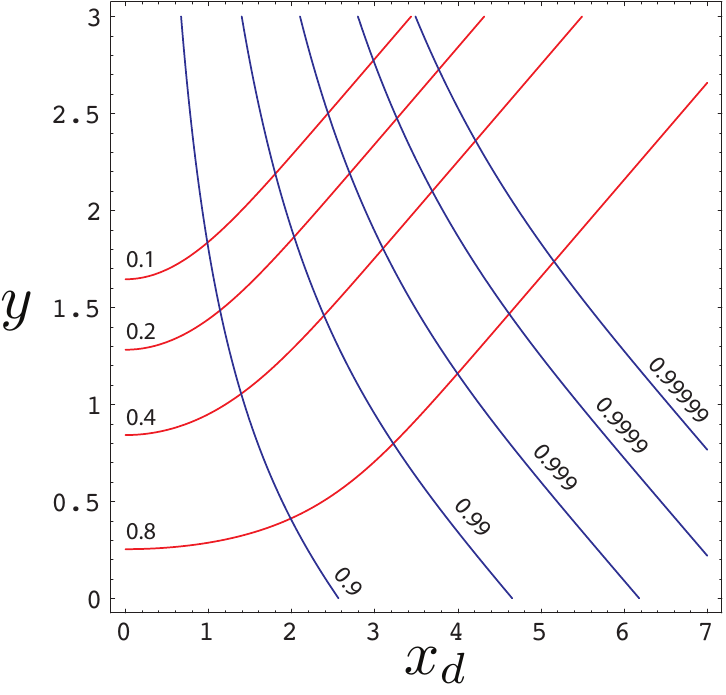}
\caption{\footnotesize Contour lines for post-selected fidelity $F_{p,y}$ of one-bit teleportation from the qubus to a qubit (blue), and success probability for post-selection (red), as a functions of $x_d$ and $y$.}
\label{onebitpostselfidelity}
\end{figure}

Instead of accepting all states corresponding to all $x$ outcomes of the homodyne measurement which implements $\aZ$, we only accept states corresponding to outcomes which are far enough away from the midpoint $x_0$, since the state at $x_0$ has the lowest fidelity with the desired state. More explicitly, we only accept states corresponding to measurement outcomes which are smaller than $x_0-y$ or larger than $x_0+y$. This post-selection can only be performed for one-bit teleportation from the qubus to the qubit, yielding a probability of success given by
\begin{multline}
\Pr(|x-x_0|>y) =\\\frac{1}{2}\left[\erfc\left(\frac{2y-x_d}{2\sqrt{2}}\right)+\erfc\left(\frac{2y+x_d}{2\sqrt{2}}\right)\right]\label{eqn:SingeTelePostP},
\end{multline}
and process fidelity conditioned on the successful outcome given by
\begin{equation}
F_{p,y} = \frac{\erfc\left(\frac{2y-x_d}{2\sqrt{2}}\right)}{\erfc\left(\frac{2y-x_d}{2\sqrt{2}}\right)+\erfc\left(\frac{2y+x_d}{2\sqrt{2}}\right)}\label{eqn:SingeTelePostF}.
\end{equation}
The effect of discarding some of the states depending on the measurement outcome for the teleportation in Fig.~(\ref{one-teleWNL}b) is depicted in Fig.~(\ref{onebitpostselfidelity}). 
In particular, we see that the process fidelity can be made arbitrarily close to $1$ at the cost of lower probability of success, while $\alpha$ and $\theta$ are unchanged, since
\begin{equation}
\lim_{y\to\infty}F_{p,y}=1.
\end{equation}
As the probability mass is highly concentrated due to the Gaussian shape of the wave packets, the probability of success drops super-exponentially fast as a function of $y$. This is because for large $z$ we have~\cite{wolfram}
\begin{equation}
\frac{2}{\sqrt{\pi}} \frac{e^{-z^2}}{{z+\sqrt{z^2+2}}}
<
\erfc(z)
<
\frac{2}{\sqrt{\pi}} \frac{e^{-z^2}}{{z+\sqrt{z^2+\frac{4}{\pi}}}}.
\end{equation}
This fast decay corresponds to the contour lines for decreasing probability of success getting closer and closer in Fig.~(\ref{onebitpostselfidelity}). Thus, while the fidelity can be increased arbitrarily via post-selection (by increasing $y$), this leads to a drop in the probability of obtaining the successful outcome for post-selection. Note that, despite this scaling, significant gains in fidelity can be obtained by post-selection while maintaining the physical resources such as $\alpha$ and $\theta$ fixed, and while maintaining a reasonable probability of success. In particular, if $x_d=2.5$, increasing $y$ from $0$ to $1.25$ takes the fidelity from $0.9$ to $0.99$ while the probability of success only drops from $1$ to $0.5$.

If the probability of success is to be maintained constant, a linear increase in $x_d$ can bring the fidelity exponentially closer to unity, as is evident in Fig.~(\ref{onebitpostselfidelity}). As $x_d$ is proportional to the amplitude $\alpha$ of the coherence state, this can be achieved while maintaining $\theta$ constant. Since $\theta$ is usually the parameter which is hard to increase in an experimental setting, this is highly advantageous.

Instead of discarding the outputs with unacceptable fidelity, one can also use the information that the failure is heralded to recover and continue the computation. In the case of the one-bit teleportations described here, such an approach would require active quantum error correction or quantum erasure codes -- the type of codes necessary for heralded errors -- which have much higher thresholds than general quantum error correcting codes~\cite{Knill}. We will not discuss such a possibility further in this paper, and will focus instead on post-selection for quantum gate construction and state preparation.

\section{\label{sec:coherentstatelogic}Universal Computation with Qubus Logic}

Previous work by Ralph {\em et al.}~\cite{Ralph02,Ralph03} and Gilchrist {\em et al.}~\cite{Gilchrist04} illustrated the construction of a universal quantum computer using what we call {\em coherent state logic}.  In these schemes a universal set of gates is applied to qubit basis states defined as $\ket{0}_{L}=\ket{-\alpha'}$ and $\ket{1}_{L}=\ket{\alpha'}$, using partial Bell state measurements and cat states of the form $\left(\ket{-\alpha'}+\ket{\alpha'}\right)/\sqrt{2}$ as resources. To perform a universal set of gates a total of sixteen ancilla cat states are necessary~\cite{Ralph03}. For $\alpha'\geq 2$ the qubits $\ket{-\alpha'}$ and $\ket{\alpha'}$ are approximately orthogonal since  $|\bk{\alpha'}{-\alpha'}|^{2}=e^{-4\alpha'^{2}}\leq 10^{-6}$. 

Using the one-bit teleportations in Fig.~(\ref{one-teleWNL}) we can also perform a universal set of gates on a Hilbert space spanned by the states $\ket{\mathbf{0}}_{L}=\ket{\alpha}$ and $\ket{\mathbf{1}}_{L}=\ket{e^{\pm i\theta}\alpha}$, which we call {\em qubus logic}. As mentioned in the previous section, the two states defined for the logical $\ket{\mathbf{1}}_{L}$ are indistinguishable when we homodyne detect along the $x$-quadrature, a fact that will become important later.  The overlap between these basis states $|\bk{\alpha}{e^{\pm i\theta}\alpha}|^{2}=e^{-2|\alpha|^{2}(\cos\theta-1)}\approx e^{-|\alpha|^{2}\theta^{2}}$ (for small $\theta$) is close to 0 provided $\alpha\theta\gg1$, so that we may consider them orthogonal -- e.g. for $\alpha\theta>3.4$, we have $|\bk{\alpha}{e^{i\theta}\alpha}|^{2}\leq 10^{-6}$. It can be seen that our basis states are equivalent to the basis states of coherent state logic given a displacement and a phase shifter. That is, if we displace the arbitrary state $a\ket{\alpha}+b\ket{\alpha e^{i\theta}}$ by $D(-\alpha\cos\left(\theta/2\right)e^{i\theta/2})$ and apply the phase shifter $e^{i(\pi-\theta)\hat{n}/2}$ we have $a\ket{\alpha\sin\left(\theta/2\right)}+be^{i\alpha^{2}\sin(\theta)/2}\ket{-\alpha\sin\left(\theta/2\right)}$. If we now set $\alpha'=\alpha\sin\left(\theta/2\right)\approx \alpha\theta/2$, for small $\theta$, we see that our arbitrary qubus logical state is equivalent to an arbitrary coherent state qubit. The $e^{i\alpha^{2}\sin(\theta)/2}$ phase factor can be corrected once we use a single bit teleportation.  If $\alpha'\geq2$ then $\alpha\theta\geq 4$, which is already satisfied by the approximate orthogonality condition $\alpha\theta\gg1$. It is important to note that, although the basis states are equivalent, the gate constructions we describe for qubus logic are very different than the gate constructions for coherent state logic. 

We compare qubus logic and coherent state logic based on resource usage, i.e. the number of ancilla states and controlled rotations necessary to perform each operation.
Since the cat state ancillas needed in coherent state logic,  $(\ket{-\alpha'}+\ket{\alpha'})/\sqrt{2}$, can be made using the circuit in Fig.~(\ref{one-teleWNL}a) with an incident photon in the state $(\ket{0}+\ket{1})/\sqrt{2}$ provided $\alpha'=\alpha\sqrt{(1-\cos(\theta))/2}\approx \alpha\theta/2$, we consider the sixteen ancilla cat states required in~\cite{Ralph03} for a universal set of gates to be equivalent to sixteen controlled rotations. 

In the next two sections, we describe how to construct arbitrarily good approximations to any single qubit unitary rotation as well the unitary $\text{CSIGN}=\text{diag}(1,1,1,-1)$ in qubus logic, as this is sufficient for universal quantum computation~\cite{DiVincenzo95}. 

\subsection{Single Qubit Gates}
\begin{figure}[ht]
\includegraphics[width=8cm]{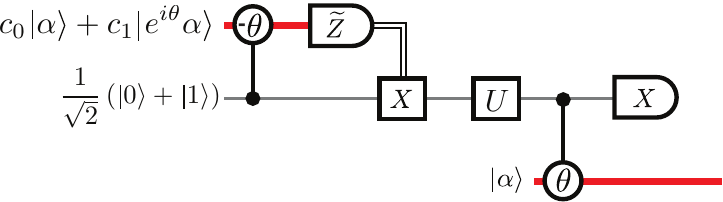}
\caption{\footnotesize A single qubit gate performed on  $c_{0}\ket{\alpha}+c_{1}\ket{e^{i\theta}\alpha}$.  } 
\label{SingleQubitGate}
\end{figure}

An arbitrary single qubit unitary gate $U$ can be applied to the state $c_{0}\ket{\alpha}+c_{1}\ket{e^{i\theta}\alpha}$ by the circuit shown in Fig.~(\ref{SingleQubitGate}). We first teleport this state to the qubit using the circuit in Fig.~(\ref{one-teleWNL}b) and then perform the desired unitary $U$ on the qubit, giving $U\bigl(c_{0}\ket{0}+c_{1}\ket{1}\bigr)$. We can teleport this state back to the qubus mode with Fig.~(\ref{one-teleWNL}a), while the $\aZ$ correction can be delayed until the next single qubit gate, where it can be implemented by applying a $Z$ in addition to the desired unitary. If it happens that this single qubit rotation is the last step of an algorithm, we know that this $\tilde{Z}$ error will not effect the outcome of a homodyne measurement (which is equivalent to a measurement in the Pauli Z eigenbasis), so that this correction may be ignored. In total this process requires two controlled rotations.
  
Since arbitrary single qubit gates are implemented directly in the two level system, the only degradation in the performance comes from the teleportation of the state from the qubus to the qubit, resulting in the fidelity given in Eq.~(\ref{eqn:SingeTeleF})

In the case that we wish to perform a bit flip on the qubit $c_{0}\ket{\alpha}+c_{1}\ket{e^{i\theta}\alpha}$ we can simply apply the phase shifter $e^{-i\theta\hat{n}}$ to obtain $c_{0}\ket{e^{-i\theta}\alpha}+c_{1}\ket{\alpha}$, similarly to the bit flip gate in~\cite{Ralph03}. 

\subsubsection{Post-selected implementation of single qubit gates}

The fidelity of single qubit gates in qubus logic can be improved simply by using post-selected teleportations. For simplicity, if we disregard the second one-bit teleportation which transfers the state back to qubus logic, we obtain the probability of success given in Eq.~(\ref{eqn:SingeTelePostP}) and the conditional process fidelity given in Eq.~(\ref{eqn:SingeTelePostF}).

\subsection{Two Qubit Gates}
To implement the entangling CSIGN gate we teleport our qubus logical state onto the polarisation entangled state $\frac{1}{2}\bigl(\ket{00}+\ket{01}+\ket{10}-\ket{11}\bigr)$. The state $\frac{1}{2}\bigl(\ket{00}+\ket{01}+\ket{10}-\ket{11}\bigr) = (\1\otimes H) (\ket{00}+\ket{11})/\sqrt{2}$, where $H$ represents a Hadamard gate, can be produced offline by any method that generates a maximally entangled pair of qubits. As described previously in the context of error correction, such a state can be produced with controlled rotations~\cite{StabPap}. If we start with the qubus coherent state $\ket{\sqrt{2}\alpha}$ and an eigenstate of the Pauli $X$ operator $(\ket{0}+\ket{1})/\sqrt{2}$ incident on Fig.~(\ref{one-teleWNL}a), we obtain $\ket{\sqrt{2}\alpha}+\ket{\sqrt{2}e^{i\theta}\alpha}$. Next we put this through a symmetric beam splitter to obtain $\frac{1}{\sqrt{2}}\bigl(\ket{\alpha,\alpha}+\ket{e^{i\theta}\alpha,e^{i\theta}\alpha}\bigr)$~\cite{Gilchrist04}. If we now teleport this state to polarisation logic with Fig.~(\ref{one-teleWNL}b) we have, to a good approximation, the Bell state $\bigl(\ket{00}+\ket{11}\bigr)/\sqrt{2}$, and with a local Hadamard gate we finally obtain $\frac{1}{2}\bigl(\ket{00}+\ket{01}+\ket{10}-\ket{11}\bigr)$. To make this state we have used three controlled rotations and one ancilla photon. Since we are only concerned with preparing a resource state which in principle can be stored, we can perform post-selection at the teleportations to ensure the state preparation is of high fidelity, as described in Section~\ref{sec:postselectedteleport}.

\begin{figure}[ht]
\includegraphics[width=8.7cm]{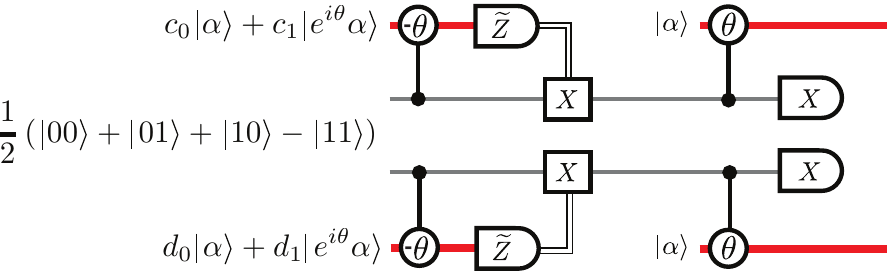}
\caption{\footnotesize Circuit used to perform a CSIGN between states in qubus logic.} 
\label{TwoQubitGate}
\end{figure}

After this gate teleportation onto qubits, we teleport back to the qubus modes after a possible $X$ correction operation. The overall circuit is shown in Fig.~(\ref{TwoQubitGate}). This CSIGN gate requires four controlled rotations. As with the single qubit gates, $\tilde{Z}$ corrections may be necessary after the final teleportations of Fig.~(\ref{TwoQubitGate}), but these corrections can also be delayed until the next single qubit gate. 

We can see what affect the condition $\alpha\theta^{2}\not\gg1$ has on the function of the gate in Fig.~(\ref{TwoQubitGate}) by looking at the process fidelity. As this gate operates on two qubits, the input state to the process we want to compare is 
\begin{multline}
\frac{1}{2}\left(\ket{0,0}\ket{\alpha,\alpha}+\ket{0,1}\ket{\alpha,\alpha e^{i\theta}}\right.\\\left.+\ket{1,0}\ket{\alpha e^{i\theta},\alpha}+\ket{1,1}\ket{\alpha e^{i\theta},\alpha e^{i\theta}}\right).
\end{multline}

\begin{figure}[htb]
\centering
\includegraphics[width=8cm]{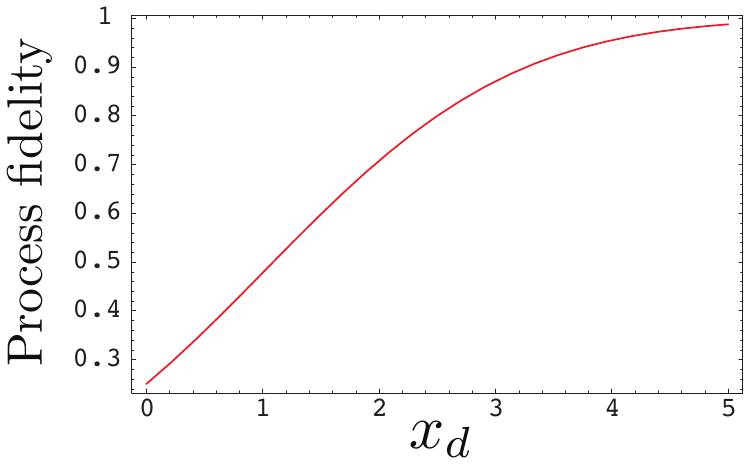}
\caption{\footnotesize Fidelity $F_{\text{CSIGN}}$ of one-bit CSIGN teleportation from the qubus to a qubit, as a function of $x_d$.}
\label{csignfidelity}
\end{figure}

From the basis states we have defined, we expect the output
\begin{multline}
\ket{\psi_2}=\frac{1}{2}\left(\ket{0,0}\ket{\alpha,\alpha}+\ket{0,1}\ket{\alpha,\alpha e^{i\theta}}\right.\\\left.+\ket{1,0}\ket{\alpha e^{i\theta},\alpha}-\ket{1,1}\ket{\alpha e^{i\theta},\alpha e^{i\theta}}\right).
\end{multline}
The unnormalised state output from Fig.~(\ref{TwoQubitGate}) is
\begin{widetext}
\begin{multline}\label{eqn:csignhomostate}
\ket{\psi_{2,o}}=\frac{1}{4}\Bigl\{
f(x,\alpha)f(x',\alpha) 
\left[\ket{00}\ket{00}+\ket{01}\ket{01}+\ket{10}\ket{10}-\ket{11}\ket{11}\right] \\
+f(x,\alpha)f(x',\alpha\cos(\theta))
\left[e^{-i\phi(x')}(\ket{00}\ket{01}+\ket{10}\ket{11})+e^{i\phi(x')}(\ket{01}\ket{00}-\ket{11}\ket{10})\right]\\
+f(x,\alpha\cos(\theta))f(x',\alpha)
\left[e^{-i\phi(x)}(\ket{00}\ket{10}+\ket{01}\ket{11})+e^{i\phi(x)}(\ket{10}\ket{00}-\ket{11}\ket{01})\right]\\
\left.+f(x,\alpha\cos(\theta))f(x',\alpha\cos(\theta))
\left[e^{-i(\phi(x)+\phi(x'))}\ket{00}\ket{11}+e^{i(\phi(x')-\phi(x))}\ket{01}\ket{10}+\right.\right.\\
\left.e^{i(\phi(x)-\phi(x'))}\ket{10}\ket{01}-e^{i(\phi(x)+\phi(x'))}\ket{11}\ket{00}\right]\Bigr\},
\end{multline}
\end{widetext}
where $x$ and $x'$ are the outcomes of the $\aZ$ measurements (top and bottom in Fig.~(\ref{TwoQubitGate}), respectively). For simplicity, we disregard the final teleportations back to qubus modes, as we have already discussed how they affect the average fidelity of the state in Section~\ref{sec:onebitteleportations}. Since we have two homodyne measurements to consider, we need to look at the four cases:  (i) $x$ greater than $x_0$ and $x'$ greater than $x_0$; (ii) $x$ greater than $x_0$ and $x'$ less than $x_0$; (iii) $x$ greater than $x_0$ and $x'$ less than $x_0$;  (iv) $x$ less than $x_0$ and $x'$ less than $x_0$. The necessary corrections for each of these cases are (i) $\1\otimes\1$ (ii) $\1\otimes Z_{\phi(x')}X$ (iii) $ Z_{\phi(x)}X \otimes\1 $ (iv) $Z_{\phi(x)}X\otimes Z_{\phi(x')}X$. Integrating over $x$ and $x'$ for these four different regions, one finds the process fidelity to be 
\begin{equation}
F_{\text{CSIGN}}=\frac{1}{4}\left(1+\text{erf}\left(\frac{x_{d}}{2\sqrt{2}}\right)\right)^{2},
\end{equation}
which just corresponds to the square of the process fidelity for a one-bit teleportation into qubits, as the only source of failure is the indistinguishability of the basis states for qubus logic. A plot showing how this fidelity scales as a function of $x_{d}$ is shown in Fig.~(\ref{csignfidelity}). 

\subsubsection{Post-selected implementation of the entangling gate}

We can counteract the reduction in fidelity shown in Fig.~(\ref{csignfidelity}) in a similar way to the single qubit gate case, by only accepting measurement outcomes less than $x_0-y$ and greater than $x_0+y$. We find the success probability and conditional fidelity to be
\begin{gather}
P_{\text{CSIGN}}=\frac{1}{4}\left(\text{erfc}\left(\frac{2y-x_{d}}{2\sqrt{2}}\right)+\text{erfc}\left(\frac{2y+x_{d}}{2\sqrt{2}}\right)\right)^{2}\\
F_{\text{CSIGN},y}=\left(\frac{\text{erfc}\left(\frac{2y-x_{d}}{2\sqrt{2}}\right)}{\text{erfc}\left(\frac{2y-x_{d}}{2\sqrt{2}}\right)+\text{erfc}\left(\frac{2y+x_{d}}{2\sqrt{2}}\right)}\right)^{2},
\end{gather}
respectively. As before, we see that the process fidelity can be made arbitrarily close to $1$ at the cost of lower probability of success. It should also be immediately clear that as $y\to0$, we have $P_{\text{CSIGN}}\to1$ and $F_{\text{CSIGN},y}\to F_{\text{CSIGN}}$. 

\begin{figure}[htb]
\centering
\includegraphics{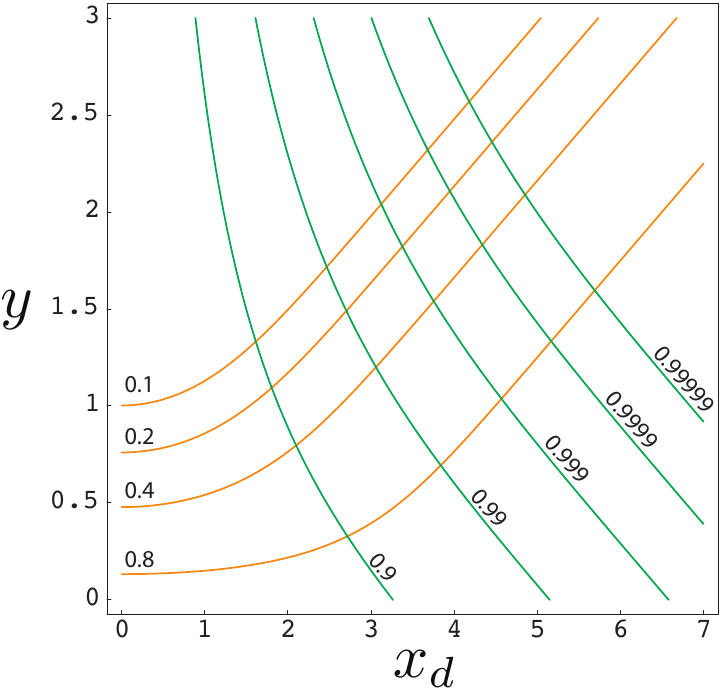}
\caption{\footnotesize Contour lines for post-selected fidelity $F_{\text{CSIGN},y}$ of CSIGN teleportation from the qubus to a qubit (green), and success probability for post-selection (gold), as a functions of $x_d$ and $y$.}
\label{csignpostselfidelity}
\end{figure}

We see the effect of ignoring some of the homodyne measurements in Fig.~(\ref{csignpostselfidelity}).  Even though performance is degraded because of the use of two one-bit teleportations, the general scalings of the fidelity and probability of success with respect to $y$ and $x_d$ are similar to the one-bit teleportation. In particular, we see that the fidelity can be increased significantly by increasing $x_d$ (or equivalently, $\alpha$).

\subsection{Comparison between Qubus Logic and Coherent State Logic}
The total number of controlled rotations necessary to construct our universal set of quantum gates on qubus logic, consisting of an arbitrary single qubit rotation and a CSIGN gate, is nine -- the construction of an arbitrary single qubit gate required two controlled rotations and the construction of  a CSIGN gate required seven, three for the entanglement production and four for the gate operation. This is in contrast to the sixteen controlled rotations (where we assume each controlled rotation is equivalent to a cat state ancilla) necessary for a universal set of gates in coherent state logic~\cite{Ralph02,Ralph03, Gilchrist04}, where an arbitrary single qubit rotation is constructed via $\exp\left(-i \frac{\vartheta }{2}Z\right)\exp\left(-i \frac{\pi }{4}X\right)\exp\left(-i \frac{\varphi }{2}Z\right)\exp\left(i \frac{\pi }{4}X\right)$, with each rotation requiring two cat state ancilla, and a CNOT gate requiring eight cat state ancilla. 

 As a further comparison we compare the resource consumption of the qubus logic scheme with the recent extension to the coherent state logic scheme by Lund~{\em et al.}~\cite{Lund:PRL} that considers small amplitude coherent states. In this scheme gate construction is via unambiguous gate teleportation, where the failure rate for each teleportation is dependent on the size of the amplitude of the coherent state logical states. Each gate teleportation requires offline probabilistic entanglement generation. On average, an arbitrary rotation about the $Z$ axis would require three cat state ancilla and both the Hadamard and CSIGN gate would each require 27 cat state ancilla. 
 
 The scheme proposed here yields significant savings compared to previous schemes in terms of the number of controlled rotations necessary to apply a universal set of gates on coherent states.

\section{\label{sec:clusterstate}Construction of Cluster States}

As we have pointed out in the previous section, the GHZ preparation scheme used for fault-tolerant error correction with strong coherent beams~\cite{StabPap} can be used to perform CSIGN gate teleportation. This approach can be generalised to aid in the construction of cluster states~\cite{Raussendorf01}, as GHZ states are locally equivalent to star graph states~\cite{Hein06, Campbell07}. Once we have GHZ states we can either use CNOT gates built with the aid of a qubus~\cite{Nemoto04,Munro05b} to deterministically join them to make a large cluster state, or use fusion gates~\cite{Browne05} to join them probabilistically. 

Recent work by Jin~{\em et al.}~\cite{Jin:PRA} showed a scheme to produce arbitrarily large cluster states with a single coherent probe beam. In this scheme, $N$ copies of the state $\left(\ket{H}+\ket{V}\right)/\sqrt{2}$ can be converted into the GHZ state $(\ket{H}^{\otimes N}+\ket{V}^{\otimes N})/\sqrt{2}$ with the use of $N$ controlled rotations and a single homodyne detection  . However, the size of the controlled rotations necessary scales exponentially with the size of the desired GHZ state -- the $N'$th controlled rotation would need to be $2^{N-1}-1$ times larger than the first controlled rotation applied to the probe beam. For example, if we consider an optimistic  controlled rotation $\theta$ of order $0.1$, once $N$ reaches 10 we would require a controlled rotation on the order of $\pi$, which is unfeasible for most physical implementations.  
In the next section we describe how to prepare GHZ states that only require large amplitude coherent states, while using the same fixed controlled rotations $\theta$ and $-\theta$.

\subsection{GHZ State Preparation and Repetition Encoding}

We mentioned a scheme in the previous section to construct the Bell state $\ket{00}+\ket{11}$, but this can be generalised to prepare GHZ states of any number of subsystems. We first start with the state $(\ket{0}+\ket{1})/\sqrt{2}$ and teleport it to a qubus initially in the larger amplitude $\ket{\sqrt{N}\alpha}$. This will give $(\ket{\sqrt{N}\alpha}+\ket{\sqrt{N}\alpha e^{i\theta}})/\sqrt{2}$. Sending this state through an $N$ port beam splitter with $N-1$ vacuum states in the other ports gives $(\ket{\alpha}^{\otimes N}+\ket{\alpha e^{i\theta}}^{\otimes N})/\sqrt{2}$. Each of these modes can then be teleported back to qubits, yielding $(\ket{0}^{\otimes N}+\ket{1}^{\otimes N})/\sqrt{2}$. The resources that we use to make a GHZ state of size $N$ are $N+1$ controlled rotations, $N+1$ single qubit ancillas, a single qubit measurement and $N$ homodyne detections.

\begin{figure}[htb]
\centering
\includegraphics[width=8cm]{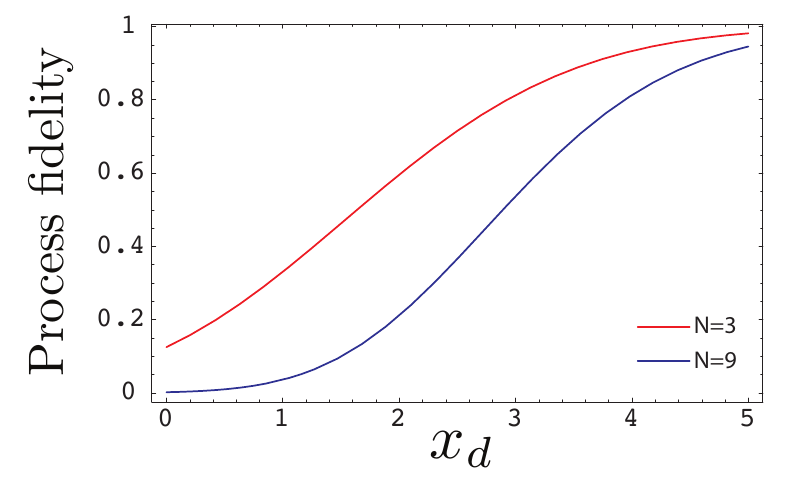}
\caption{\footnotesize Process fidelity $F_{\text{REP}}$ of repetition encoding as a function of $x_d$.}
\label{repfidelity}
\end{figure}

This circuit can also function as an encoder for a quantum repetition code, in which case we can allow any input qubit state $a\ket{0}+b\ket{1}$ and obtain an approximation to $a\ket{0}^{\otimes N}+b\ket{1}^{\otimes N}$. In order to evaluate the performance of this process, we once again calculate the process fidelity by using the input state $\frac{1}{\sqrt{2}}(\ket{00}+\ket{11})$ and acting on the second subsystem. Using a generalisation of Eqn.~(\ref{eqn:csignhomostate}) we calculate the effect of $\alpha\theta^{2}\not\gg1$ on the production of a GHZ state of size $N$ to be 
\begin{equation}\label{repfidel}
F_{\text{REP}}=\frac{1}{2^{N}}\left(1+\text{erf}\left(\frac{x_{d}}{2\sqrt{2}}\right)\right)^{N}.
\end{equation}
Again, this corresponds to the process fidelity of a single one-bit teleportation into a qubit raised to the $N$th power. The fidelity of preparing repetition encoded states drops exponentially in $N$. In Fig.~(\ref{repfidelity}) we show the fidelity as a function of $x_{d}$ for $N=3$ and for $N=9$. 

\subsection{Post-selected Implementation of GHZ State Preparation and Repetition Encoding}
The reduction in fidelity due to $\alpha\theta^{2}\not\gg1$ in Eq.~(\ref{repfidel}) can be counteracted, as before, by simply performing post-selection during the one-bit teleportations into the qubits.

We find the success probability and conditional fidelity to be
\begin{gather}
P_{\text{REP}}=\frac{1}{2^{N}}\left(\text{erfc}\left(\frac{2y-x_{d}}{2\sqrt{2}}\right)+\text{erfc}\left(\frac{2y+x_{d}}{2\sqrt{2}}\right)\right)^{N}\\
F_{\text{REP},y}=\left(\frac{\text{erfc}\left(\frac{2y-x_{d}}{2\sqrt{2}}\right)}{\text{erfc}\left(\frac{2y-x_{d}}{2\sqrt{2}}\right)+\text{erfc}\left(\frac{2y+x_{d}}{2\sqrt{2}}\right)}\right)^{N}
\end{gather}
As $y\to0$ we see that $P_{\text{REP}}\to1$ and $F_{\text{REP},y}\to F_{\text{REP}}$. 

\begin{figure}[htb]
\centering
\includegraphics[width=8cm]{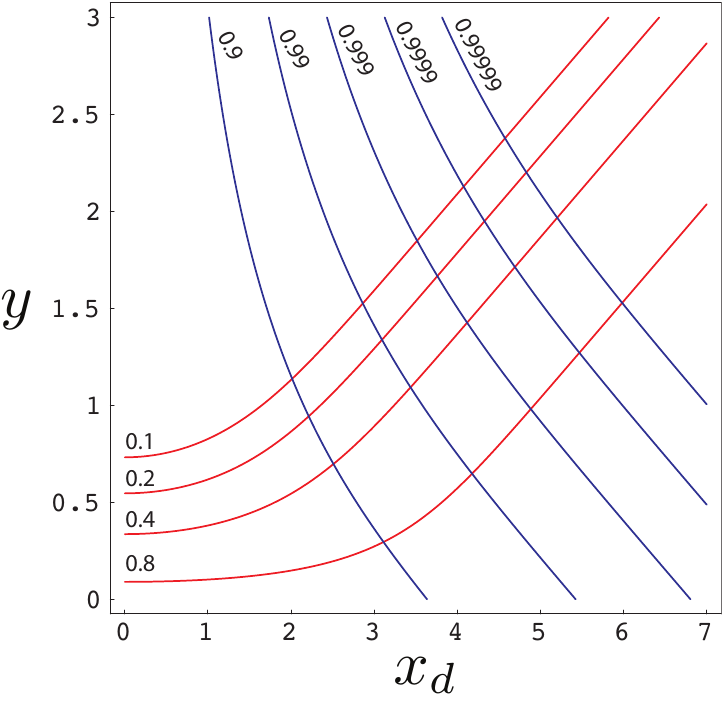}
\caption{\footnotesize Contour lines for post-selected process fidelity $F_{\text{REP},y}$ of 3-fold repetition encoding (blue), and success probability for post-selection (red), as a functions of $\alpha\theta^2$ and $y$.}
\label{reppostselfidelity}
\end{figure}

\begin{figure}[htb]
\centering
\includegraphics[width=8cm]{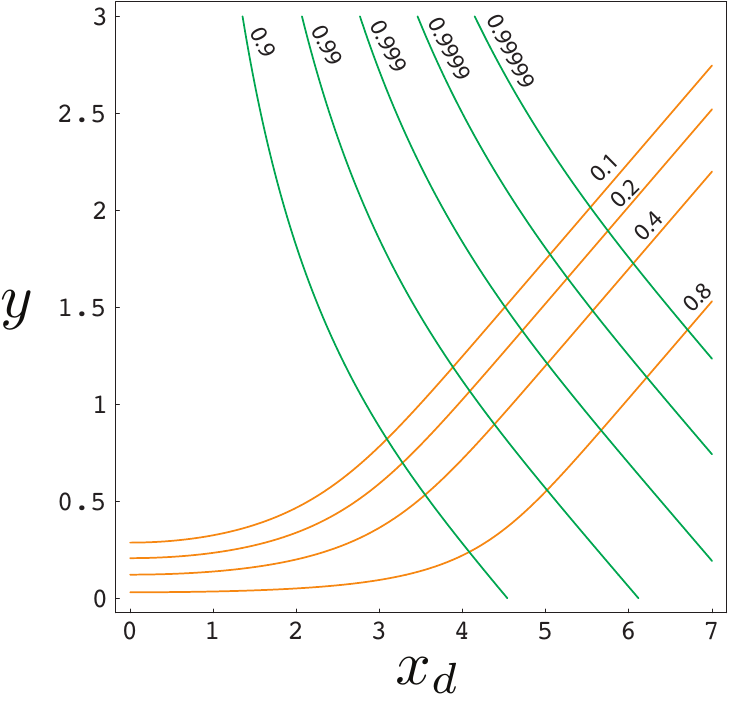}
\caption{\footnotesize Contour lines for post-selected process fidelity $F_{\text{REP},y}$ of 9-fold repetition encoding (green), and success probability for post-selection (gold), as a functions of $\alpha\theta^2$ and $y$.}
\label{reppostselfidelity9}
\end{figure}

The effect of discarding some of states corresponding to undesired homodyne measurement outcomes can be seen in Figs.~(\ref{reppostselfidelity}) and (\ref{reppostselfidelity9}). Thus,
as discussed in Section~\ref{sec:postselectedteleport}, one can prepare a state encoded in the repetition code with an arbitrarily high process fidelity, regardless of what $\theta$ and $\alpha$ are. The expected degradation in performance due to the additional teleportations is also evident in the faster decay of the probability of success with larger $y$.

\section{\label{sec:conclusions}Discussion}
We have described in detail various uses for one-bit teleportations between a qubit and a qubus. Using these teleportations, we proposed a scheme for universal quantum computation, called qubus logic, which is a significant improvement over other proposals for quantum computation using coherent states. This scheme uses fewer interactions to perform the gates, and also allows for the use of post-selection to arbitrarily increase the fidelity of the gates given any interaction strength at the cost of lower success probabilities.

The one-bit teleportations also allow for the preparation of highly entangled $N$ party states known as GHZ states, which can be used in the preparation of cluster states. Moreover, the same circuitry can be used to encode states in the repetition code which is a building block for Shor's 9 qubit code. In this case, where we are interested in preparing resource states, the power and flexibility of post-selected teleportations can be fully exploited, as the achievable fidelity of the state preparation is independent of the interaction strength available. 

The main property of the qubus which is exploited in the schemes described here is the fact that entanglement can be easily created in the qubus through the use of a beam splitter. Local operations, on the other hand, are easier to perform on a qubit. The controlled rotations allow for information to be transferred from one system to the other, allowing for the advantages of each physical system to be exploited to maximal advantage.

\begin{figure}[htb]
\centering
\includegraphics[width=8cm]{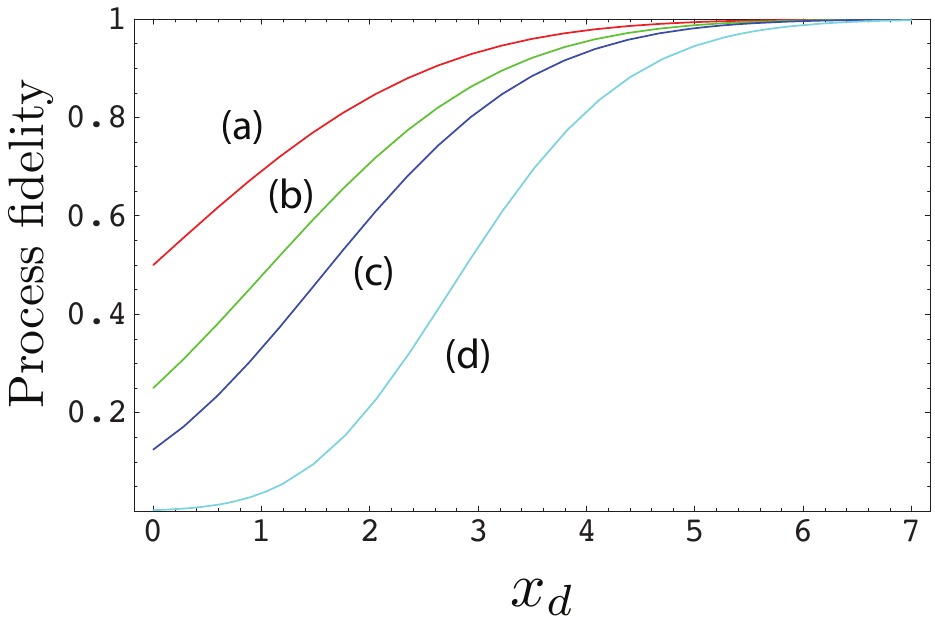}
\caption{\footnotesize Process fidelity as a function of $x_{d}$ for (a) the qubus logic single qubit gate ($F_{p}$); (b) the CSIGN teleportation ($F_{\text{CSIGN}}$); (c) repetition encoding with $N=3$ shown in blue ($F_{\text{REP}}$); (d) repetition encoding with $N=9$ ($F_{\text{REP}}$).}
\label{FidelityAllSave}
\end{figure}

\begin{figure}[htb]
\centering
\includegraphics{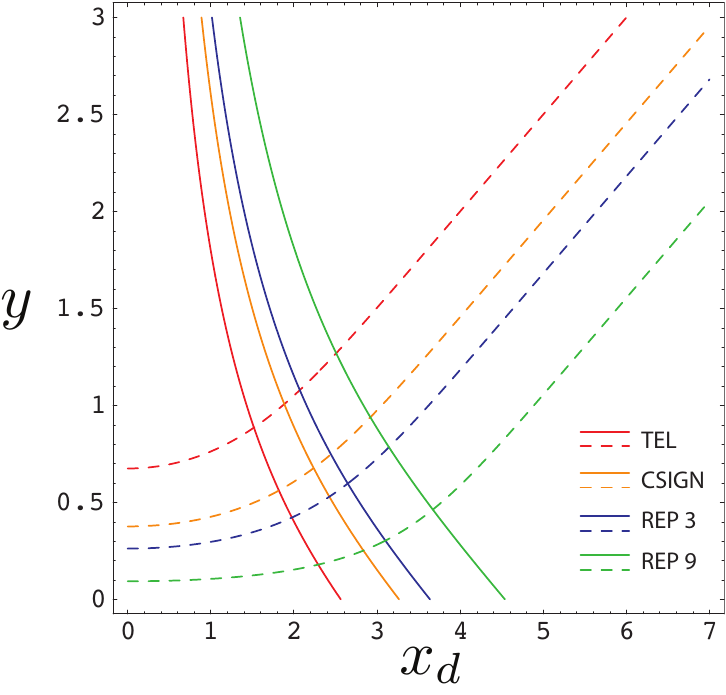}
\caption{\footnotesize Contour plot showing the conditional process fidelity (solid curves) as a function of $x_{d}$ and $y$ for $F=0.9$ for qubus to qubit one-bit teleportation (red), CSIGN teleportation (gold), repetition encoding for $N=3$ (blue) and repetition encoding for $N=9$(green). The dashed curves are contour curves for the probability of success for post-selection with $\Pr(|x-x_0|>y)=0.5$.}
\label{AllContourPlotA}
\end{figure}

The fidelity suffers as the operations become more complex, as can be seen in Figs.~(\ref{FidelityAllSave}) and (\ref{AllContourPlotA}). This is because multiple uses of the imperfect one-bit teleportation from qubus to qubit are used. As the process fidelity is less than perfect, error correction would have to be used for scalable computation. However, as we have discussed, the fact that the homodyne measurements essentially herald the fidelity of the operations, it is possible to use post-selection in conjunction with error heralding to optimise the use of physical resources.

While the scheme presented has been abstracted from particular physical implementations, any physical realisations of a qubit and a continuous variable mode would suffice. The only requirements are controlled rotations, along with fast single qubit gates and homodyne detection, which are necessary to enable feed-forward of results for the implementation of the relevant corrections. 
\begin{acknowledgments}
  We would like to thank T.C.~Ralph, K.~Nemoto and W.J.~Munro for valuable discussions. We are supported in part by NSERC, ARO, CIAR, MITACS and MEXT in Japan. C.R.M. would like to thank Mike and Ophelia Lazaridis for financial support while a student at IQC. M.S. would like to thank the Bell family for financial support.

\end{acknowledgments}

\newpage



\begin{thebibliography}{99}

\bibitem{Bennett93} C.H.~Bennett, {\em et~al.},  Phys. Rev. Lett. {\bf 70}, 1895  (1993).

\bibitem{Gottesman99} D.~Gottesman and I.L.~Chuang, Nature {\bf 402}, 390 (1999).

\bibitem{KLM} E. Knill, R. Laflamme and G.J. Milburn, Nature  {\bf 409}, 46 (2001).

\bibitem{Zhou00} X. Zhou, D.W. Leung and I.L. Chuang, Phys. Rev. A {\bf 62}, 052316 (2000).

\bibitem{StabPap} C.R. Myers, M. Silva, K. Nemoto and W.J. Munro, Phys. Rev. A {\bf 76}, 012303 (2007).

\bibitem{Raussendorf01} R.~Raussendorf and H.~J. Briegel, Phys. Rev. Lett. {\bf 86}, 5188--5191 (2001).

\bibitem{Nielsen04} M. A. Nielsen, Phys. Rev. Lett. {\bf 93}, 040503 (2004).

\bibitem{Browne05} D.E. Brown and T. Rudolph, Phys. Rev. Lett. {\bf 95}, 010501 (2005). 

\bibitem{Knill} E. Knill, Phys. Rev. A {\bf 71}, 042322 (2005). E. Knill, Nature {\bf 434}, 39-44 (2005).

\bibitem{SDKO} M. Silva, V. Danos, E. Kashefi and H. Olivier, New J. Phys. {\bf 9}, 192 (2007).

\bibitem{Munro05b}  W.J. Munro, K. Nemoto, T.P. Spiller, New J. Phys.  {\bf 7}, 137 (2005).

\bibitem{Ralph02} T.C. Ralph, W.J. Munro and G.J. Milburn, Proc. SPIE {\bf 4917}, 1 (2002); e-print quant-ph/0110115.

\bibitem{Ralph03} T.C. Ralph, A. Gilchrist, G.J. Milburn, W.J. Munro and S. Glancy, Phys. Rev. A {\bf 68}, 042319 (2003).

\bibitem{Gilchrist04} A. Gilchrist, K. Nemoto, W. J. Munro, T. C. Ralph, S. Glancy, S. L. Braunstein, G. J. Milburn, J. Opt. B:  Quantum Semiclass. Opt. {\bf 6}, S828 (2004).

\bibitem{Gardiner} C.W. Gardiner and P. Zoller, \textit{Quantum Noise}, Springer-Verlag, Berlin, (2004).

\bibitem{Barret05} S.D. Barret, P. Kok, K. Nemoto, R.G. Beausoleil, W.J. Munro and T.P. Spiller, Phys. Rev. A {\bf 71}, 060302(R) (2005).

\bibitem{Nemoto04} K. Nemoto and W.J. Munro, Phys. Rev. Lett. {\bf 93}, 250502 (2004).

\bibitem{Jamiolkowski72} A.~Jamio{\l}kowski, Rep. Math. Phys. {\bf 3}, 275 (1972).

\bibitem{Horodecki99} M.~Horodecki, P.~Horodecki and R.~Horodecki, Phys. Rev. A {\bf 60}, 1888 (1999).

\bibitem{Gilchrist05} A.~Gilchrist and N.K.~Langford and M.A.~Nielsen, Phys. Rev. A {\bf 71}, 062310 (2005).

\bibitem{wolfram} E.~W.~Weisstein, ``Erfc.'' From MathWorld -- A Wolfram Web Resource. http://mathworld.wolfram.com/Erfc.html.

\bibitem{DiVincenzo95} D.P.~DiVincenzo, Phys. Rev. A {\bf 51}, 1015 (1995).

\bibitem{Lund:PRL} A.P.~Lund, T.C.~Ralph and H.L.~Haselgrove, Phys. Rev. Lett. {\bf 100}, 030503 (2008).

\bibitem{Hein06} M.~Hein, W.~Dur, J.~Eisert, R.~Raussendorf, M.~Van den Nest and H.-J.~Briegel,  e-print quant-ph/0602096.

\bibitem{Campbell07} E.T.~Campbell, J.~Fitzsimons, S.C.~Benjamin, P.~Kok, e-print quant-ph/0702209.

\bibitem{Jin:PRA} G.-S.~Jin , Y.~Lin and B.~Wu, Phys. Rev. A {\bf 75}, 054302 (2007).


\end{thebibliography}
\end{document}